\documentclass[preprint,showpacs,preprintnumbers,amsmath,amssymb]{revtex4}
\usepackage{amssymb}
\usepackage{txfonts}

\usepackage{graphicx}
\usepackage{dcolumn}
\usepackage{bm}
\usepackage{mathrsfs}

\begin{document}

\title{From Bardeen-boson stars to  black holes without event horizon}

\author{X. E.  Wang}\email{38223000@qq.com}

\noindent \emph{This paper is dedicated to the memory of Yi-Shi Duan\\
\;\\
\;\\
}

\begin{abstract}
In a talk given in 2013, S. Hawking conjectured that the event horizon of black holes does not exist and suggested redefining black holes as bound states of the gravitational field. Inspired by this idea, we investigated the coupling of the Bardeen action and a complex scalar field model. Numerically, we obtained a class of   boson stars solutions with the magnetic monopole charge $q$.
When the constant $q$ exceeds a certain threshold, we observed that as the frequency approaches zero, a critical position $r_c$ emerges where the scalar field concentrates within its interior. Outside this critical position,
 these boson star solutions tend to infinitely approach what is known as an extreme black hole. However, there is no event horizon present.
 While our results are model-dependent and their generality remains uncertain, they align well with Hawking's conjecture that real, regular black holes do not have an event horizon and
 provided valuable insights into the understanding and development of concepts such as fuzzballs,   firewalls and black hole soft hairs.

\end{abstract}

\maketitle

\section{Introduction}
The existence of black holes is an important prediction in general relativity. Since the discovery of the first black hole solution, known as the Schwarzschild black hole, extensive research has been devoted to studying their solutions and properties, garnering significant attention.
In particular, both the discovery of gravitational waves from the merger of two BHs~\cite{LIGOScientific:2016aoc} and
the imaging of black hole shadow by
Event Horizon Telescope (EHT)~\cite{Collaboration2019a,Collaboration2019,Collaboration2019b,
Collaboration2019c,Collaboration2019d,Collaboration2019e} have further solidified the belief in the existence of black holes. However, a fundamental flaw of black hole solutions is the presence of spacetime singularities within their interiors. A.
Einstein  questioned the existence of these singularities, considering that matter cannot be infinitely compressed \cite{Einstein:1939ms}.
Currently, it is widely accepted that general relativity represents a low-energy effective theory of quantum gravity, and a complete theory of quantum gravity is expected to prevent the formation of singularities.  In addition, an important feature of a stable black hole is the existence of an event horizon, which is a  a unidirectional membrane in the classical gravitational background, allowing causal influences to cross it in only one direction, that is,  matter can only enter the event horizon but cannot escape from it.
However, a breakthrough study was discovered in 1970s \cite{Hawking:1975vcx}. By applying the principles of quantum mechanics to black holes, S. Hawking proposed that they emit a form of thermal radiation, which has since been referred to as Hawking radiation. Due to the radiation being only related to the three hairs of a black hole, it gives rise to the black hole information paradox \cite{Hawking:1976ra}. To address the information loss paradox, various proposals  have been proposed, such as the fuzzball theory \cite{Mathur:2009hf},  black hole firewalls \cite{Almheiri:2012rt} and black hole soft hair \cite{Hawking:2016msc}.
Current research consistently reveals that black holes possess internal structures within their horizons. These structures are inherently related to the distribution of matter inside them.

In a talk given in 2013, S. Hawking conjectured that the event horizon of black holes does not exist and suggested redefining black holes as bound states of the gravitational field \cite{Hawking:2014tga}. Additionally, it was proposed \cite{Hawking:2015qqa} that information is not stored within the interior of the black hole, as conventionally assumed, but rather on its boundary, known as the event horizon. Inspired by Hawking's ideas, we have considering the coupling of an  complex scalar field to  the model of Bardeen spacetime,    in which the first regular exact black hole solution  is presented \cite{Bardeen1,Ayon-Beato:1998hmi},   to investigate how  the distribution of the complex scalar fields inside and outside the event horizon.

By solving the Einstein equations, we find only one type of solution that does not possess an event horizon. These solutions are essentially a type of boson stars with magnetic charge. We refer to these solutions as Bardeen-boson stars. Boson stars is a theoretical concept in astrophysics, which are characterized by their ability to maintain stability and avoid gravitational collapse due to the balance between the attraction of gravity and the dispersive nature of the
scalar field \cite{Liebling:2012fv}.
It is worth noting that no solutions with horizons have been found. However, analyzing these solutions, we  discover that within certain frequency ranges, the scalar field of these solutions converges at a certain radius $r_c$ in radial coordinate while rapidly decaying beyond that radius, forming a steep surface.
The component  $g_{tt}$ of the metric also approaches zero rapidly due to the influence of the matter field. As the frequency approaches zero, the $g_{tt}$ is nearly zero inside
 the region  $r<r_c$.
Comparing this to the Bardeen black hole, it is observed that the functions outside this position almost degenerate into an extreme black hole. However, it is important to note that this extreme black hole is not equivalent to the extreme Bardeen black hole. From the perspective of an observer at infinity, it is impossible to distinguish whether it is an equivalent Bardeen solution or an extreme black hole.
Moreover, we calculate the energy density $\rho$ of the scalar matter field,
and find that there exists a very steep wall-liked distribution in the vicinity of the inner region near $r_c$.

The paper is organized as follows.  In Section. \ref{model}, we introduce the model of Einstein-Klein-Gordon theory coupled to a nonlinear electrodynamics.
 In Section \ref{sec3}, we present numerical results of Bardeen-boson star and analyze their physical properties. The conclusion and discussion
are given in the last section.

\section{The Model}\label{model}
In this section, we will give a brief introduction to the model of Einstein-Klein-Gordon theory coupled to a nonlinear electrodynamics. The bulk action reads
\begin{equation}\label{action}
  S=\int\sqrt{-g}d^4x\left(\frac{R}{4}+\mathcal{L}^{(1)}+\mathcal{L}^{(2)}\right),
\end{equation}
with
\begin{eqnarray}
\mathcal{L}^{(1)} &= &- \frac{ 3}{ 2 s } \left( \frac{ \sqrt{2 q^2 {\cal F}}}{ 1 + \sqrt{ 2 q^2 {\cal F}}} \right)^{\frac{5}{2}},\\
\mathcal{L}^{(2)} &= & -\nabla_a\psi^*\nabla^a\psi  - \mu^2\psi\psi^*,
\end{eqnarray}
where  $R$ is the scalar curvature, and  $\mathcal{L}^{(1)}$ is a function of ${\cal F} = \frac{1}{4}F_{ab} F^{ab}$ with the electromagnetic field strength $ F_{ab} =  \partial_{a} A_{ b} - \partial_{b} A_{ a}$, in which $A$ is  the electromagnetic field.
Here, $\psi$ is the complex scalar field. The constants $q$, $s$ and $\mu$ are three independent parameters, where $q$ represents the magnetic charge and $\mu$  represents the scalar field mass.
 By varying the action (1) with
respect to the metric, electromagnetic field and scalar field, respectively,  we can obtain the following equations of motion

\begin{eqnarray} \label{eq:EKG1}
R_{ab}-\frac{1}{2}g_{ab}R-2 (T^{(1)}_{ab}+T^{(2)}_{ab})&=&0 \ ,  \nonumber\\
\bigtriangledown_{a} \left(\frac{ \partial {\cal L}^{(1)}}{ \partial {\cal F}}  F^{a b}\right) &=& 0,    \\
\Box\psi-\mu^2\psi &=& 0, \nonumber
\end{eqnarray}
with
\begin{equation}
T^{(1)}_{ab} =- \frac{ \partial {\cal L}^{(1)}}{ \partial {\cal F}} F_{a c} F_{ b }^{\;\;c} + g_{ab} {\cal L}^{(1)},
\end{equation}
\begin{equation}
T^{(2)}_{ab} = \partial_a\psi^*\partial_b\psi + \partial_b\psi^*\partial_a\psi - g_{ab}\left[\frac{1}{2}g^{ab}\left(\partial_a\psi^*\partial_b\psi + \partial_b\psi^*\partial_a\psi\right) + \mu^2\psi^*\psi\right]\,.
\end{equation}

According to Noether's theorem, the invariance of the action of a complex scalar field under the $U(1)$ transformation $\psi\rightarrow e^{i\alpha}\psi$, where $\alpha$ is a constant, gives rise to a conserved current associated with the complex scalar field:
\begin{equation}\label{equ8}
	J^{a} = -i\left(\psi^*\partial^a\psi - \psi\partial^a\psi^*\right).
\end{equation}
By integrating the timelike component of the above conserved current over a spacelike hypersurface $\varSigma$, we can determine the Noether charge associated with the complex scalar field
\begin{equation}\label{equ9}
	Q = \frac{1}{4\pi}\int_{\varSigma}J^t .
\end{equation}
We consider the general static spherically symmetric solution, and adopt the ansatz as follows
\begin{equation}\label{equ10}
	ds^2 = -n(r)o^2(r)dt^2 + \frac{dr^2}{n(r)} + r^2\left(d\theta^2 + \sin^2\theta d\varphi^2\right),
\end{equation}
where the functions $n(r)$ and $o(r)$ only depend on the radial variable $r$. In addition, we use the following ansatzes of electromagnetic field and the scalar field:
\begin{equation}\label{equ11}
   A= q \cos(\theta)d\varphi,\;\;\; \psi = \phi(r)e^{-i\omega t},
\end{equation}

Substituting the above ansatzes (\ref{equ10}) and (\ref{equ11}) into the field equations~(\ref{eq:EKG1}), we can get the following equations
for $\phi(r)$,  $n(r)$ and $o(r)$:
\begin{eqnarray}\label{ode1}
	 \phi^{\prime\prime}+\left(\frac{2}{r} + \frac{n^\prime}{n} + \frac{o^\prime}{o}\right)\phi^\prime + \left(\frac{\omega^2}{n o^2} - \mu^2\right)\frac{\phi}{n} &=& 0\, ,\nonumber\\
 n'+2 \mu^2 r \phi^2+\frac{2 r \text{$\omega $}^2 \phi^2}{n o^2}+2 r n \phi'^2+\frac{n}{r}+\frac{3 q^5 r}{s \left(q^2+r^2\right)^{5/2}}-\frac{1}{r}&=&0,
 \\
  	\frac{o^\prime}{o} - 2r\left(\phi^{\prime2} + \frac{\omega^2\phi^2}{n^2 o^2}\right)&=&0. \nonumber
\end{eqnarray}
The Noether charges obtained from  Eq.~(\ref{equ9}) are written as
\begin{equation}\label{equ18}
	Q = 2\int_0^\infty r^2\frac{\omega\phi^2}{o~n}dr\, .
\end{equation}
Moreover, 	with the ansatz of electromagnetic field in Eq.(\ref{equ11})
the magnetic field is given by
\begin{equation}
F_{\theta \varphi } =  2 q \sin \theta,
\end{equation}
thus, we can obtain   the magnetic charge  of the magnetic monopole
\begin{equation}\label{equ18}
	\frac{1}{4\pi}\oint_{S^\infty}d A =q  .
\end{equation}

To solve the ordinary differential equations (\ref{ode1}), it is imperative to establish appropriate boundary conditions for each unknown function. Considering the asymptotically flat nature of the solutions, the metric functions $n(r)$ and $o(r)$ must conform to the following boundary conditions:
\begin{equation}\label{equ19}
n(0) = 1, \qquad o(0) = o_0, \qquad n(\infty) = 1-\frac{2 M}{r}, \qquad o(\infty) = 1,
\end{equation}
where  $o_0$  and mass $M$ of the solution are currently unknown, the values of these two quantities  can be determined  by solving  the differential equation system. Additionally,
for the complex scalar field,  we require the following boundary conditions
\begin{equation}\label{equ20}
\phi(\infty) = 0,\;\;\;\left. \frac{d\phi(r)}{dr}\right|_{r = 0} = 0.
\end{equation}

Prior to numerically solving Eq. (\ref{ode1}), it is worthwhile to investigate two special solution
cases. Firstly, for $q = 0$,   the action (\ref{action}) degenerates to Einstein-scalar theory,                the solution
which can describe the asymptotically spherically soliton, is the well-known boson star.
For the vanishing complex scalar field with $q \neq 0$, the model is the
Bardeen theory,
the solution
which can describe the asymptotically spherically soliton and black holes, is the  Bardeen spacetime, and the metric    is the
following form
\begin{equation}
ds^2 = -f(r) dt^2 + f(r)^{-1} dr^2 + r^2 ( d \theta^2 + \sin^2(\theta) d \varphi^2),
\end{equation}
where
\begin{equation}
f(r) = 1 - \frac{2M r^2}{ ( r^2 + q^2 ) ^{3/2} },
\end{equation}
with $M=q^3/2 s$.  The function $f(r)$ exhibits a local minimum at $r=\sqrt 2  q$.
 In the regime where $q<3^{3/4}\sqrt{\frac{ s}{2}}$, no horizons are present. When $q=3^{3/4}\sqrt{\frac{ s}{2}}$,  degenerate horizons are observed. Conversely, for $q>3^{3/4}\sqrt{\frac{ s}{2}}$, two distinct horizons exist.

\section{Numerical results}\label{sec3}
\begin{figure}
  \begin{center}
   \includegraphics[width=7.8cm]{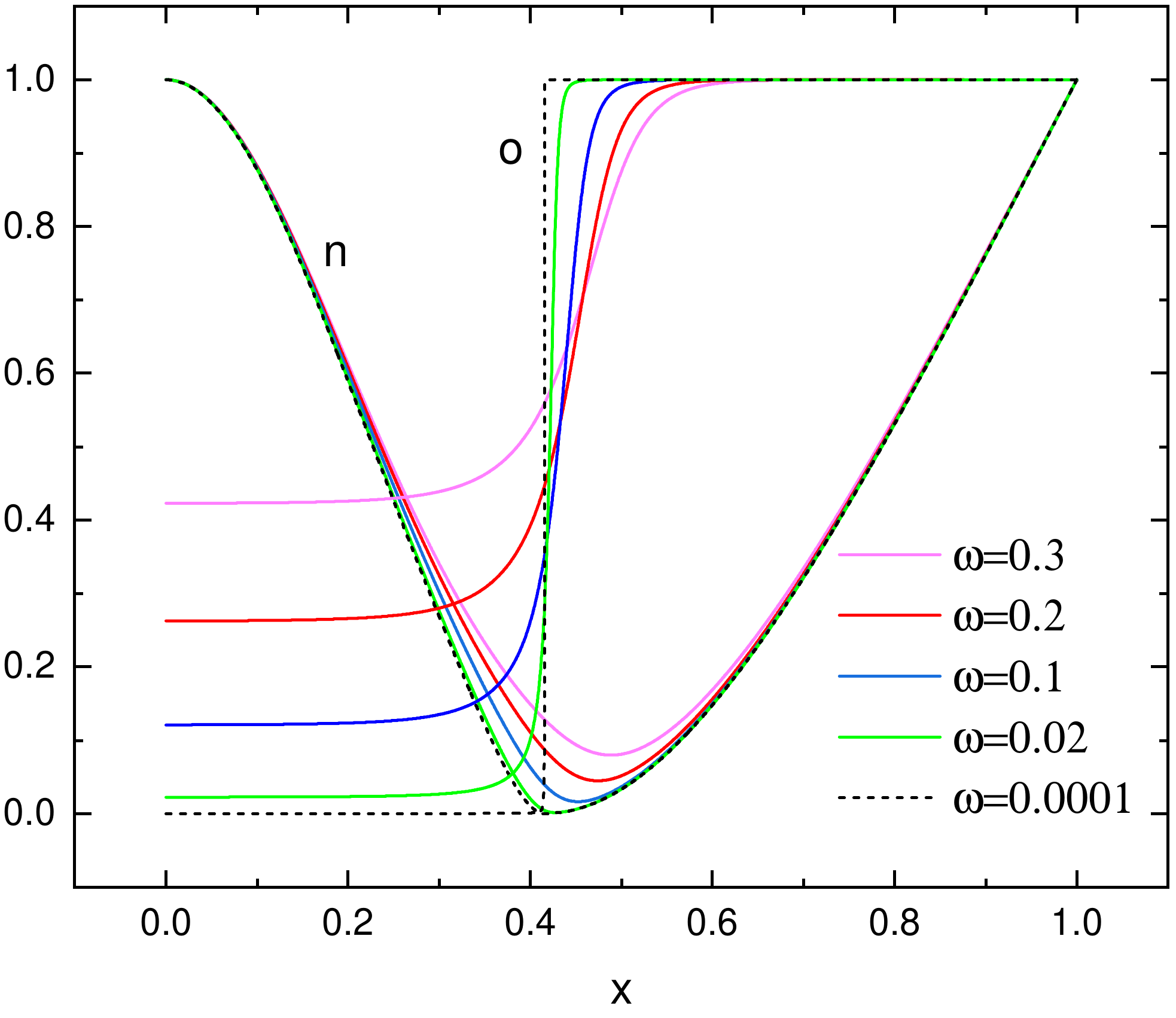}
      \includegraphics[width=8.2cm]{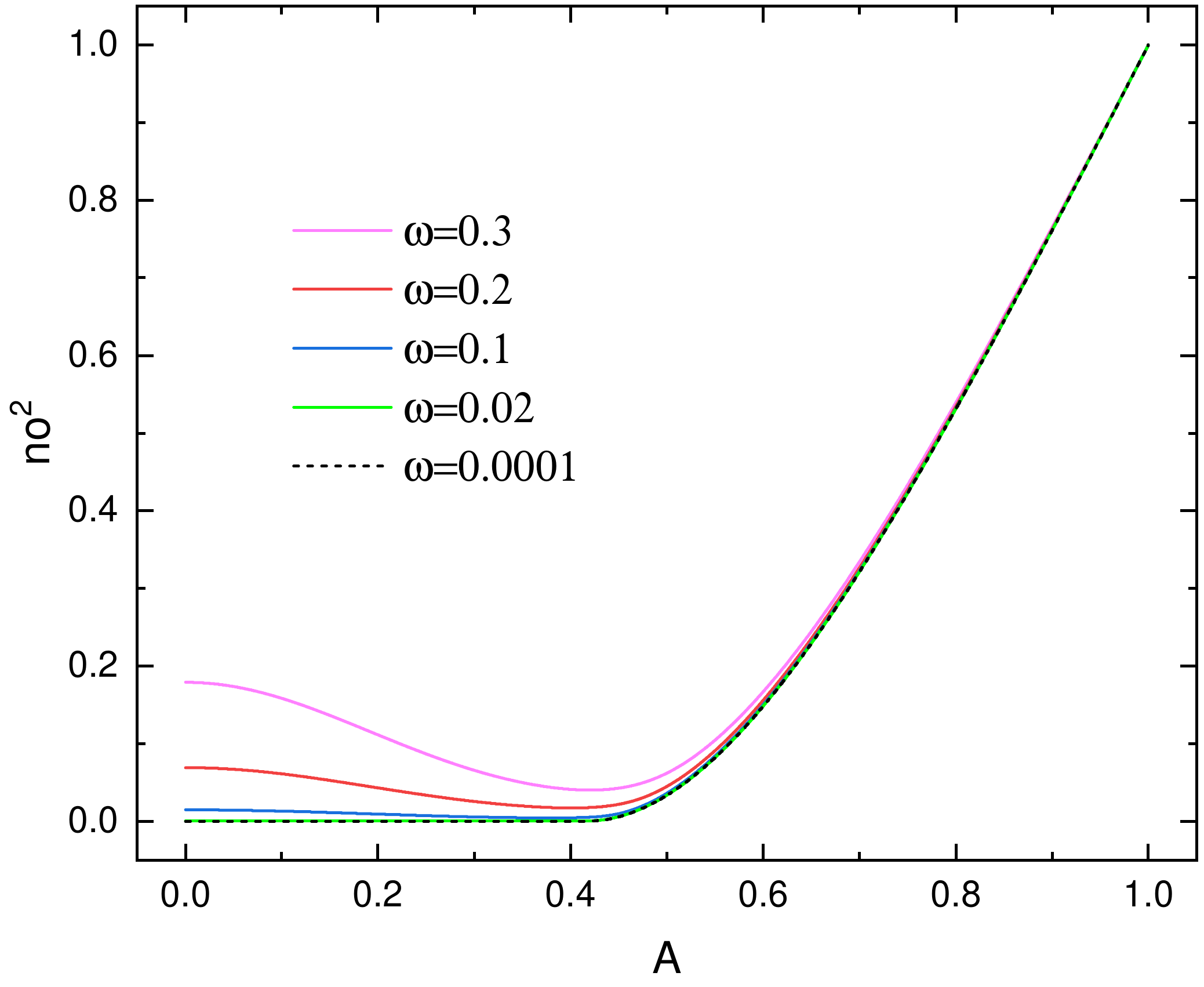}
     \includegraphics[width=8cm]{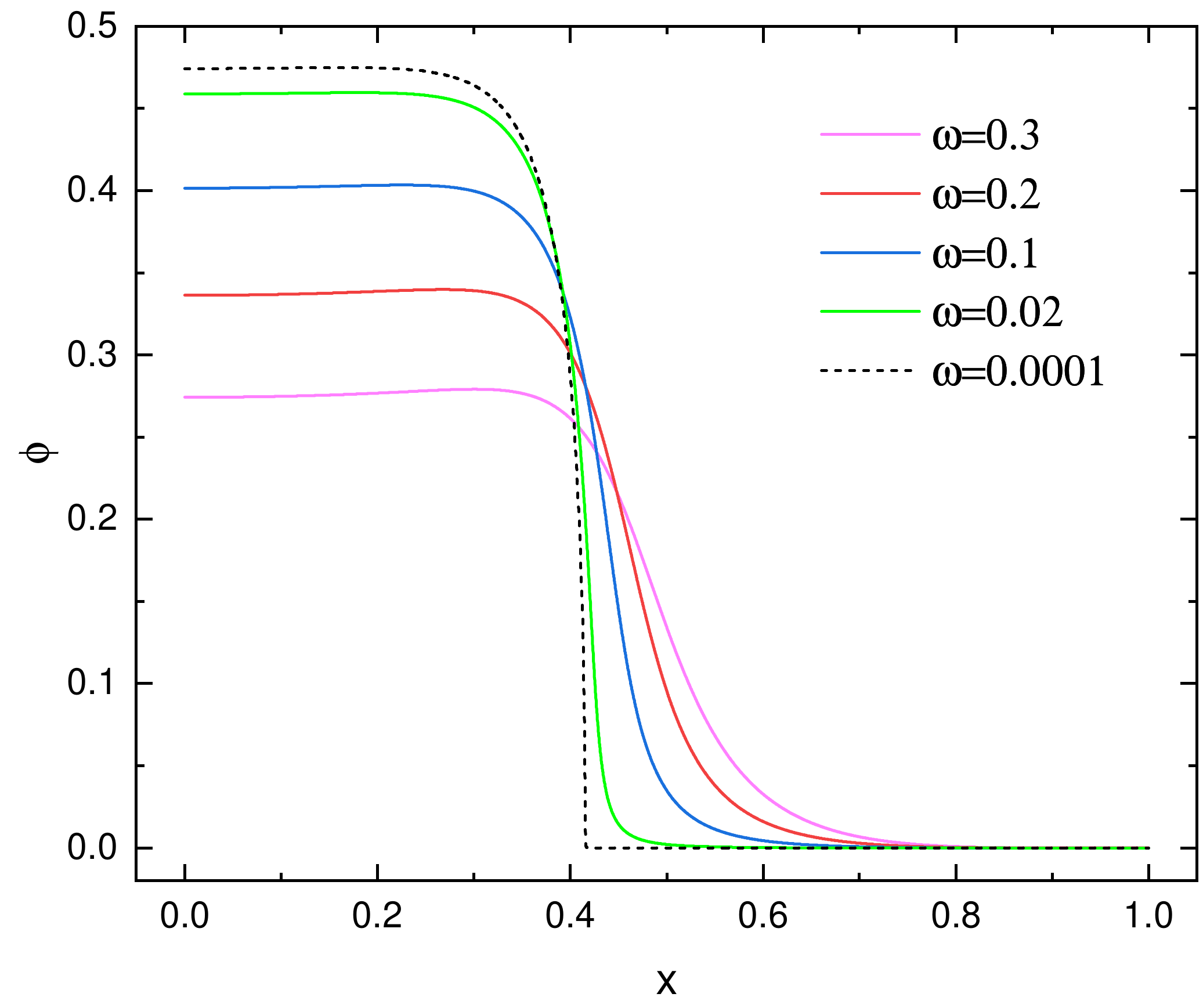}
   \includegraphics[width=8cm]{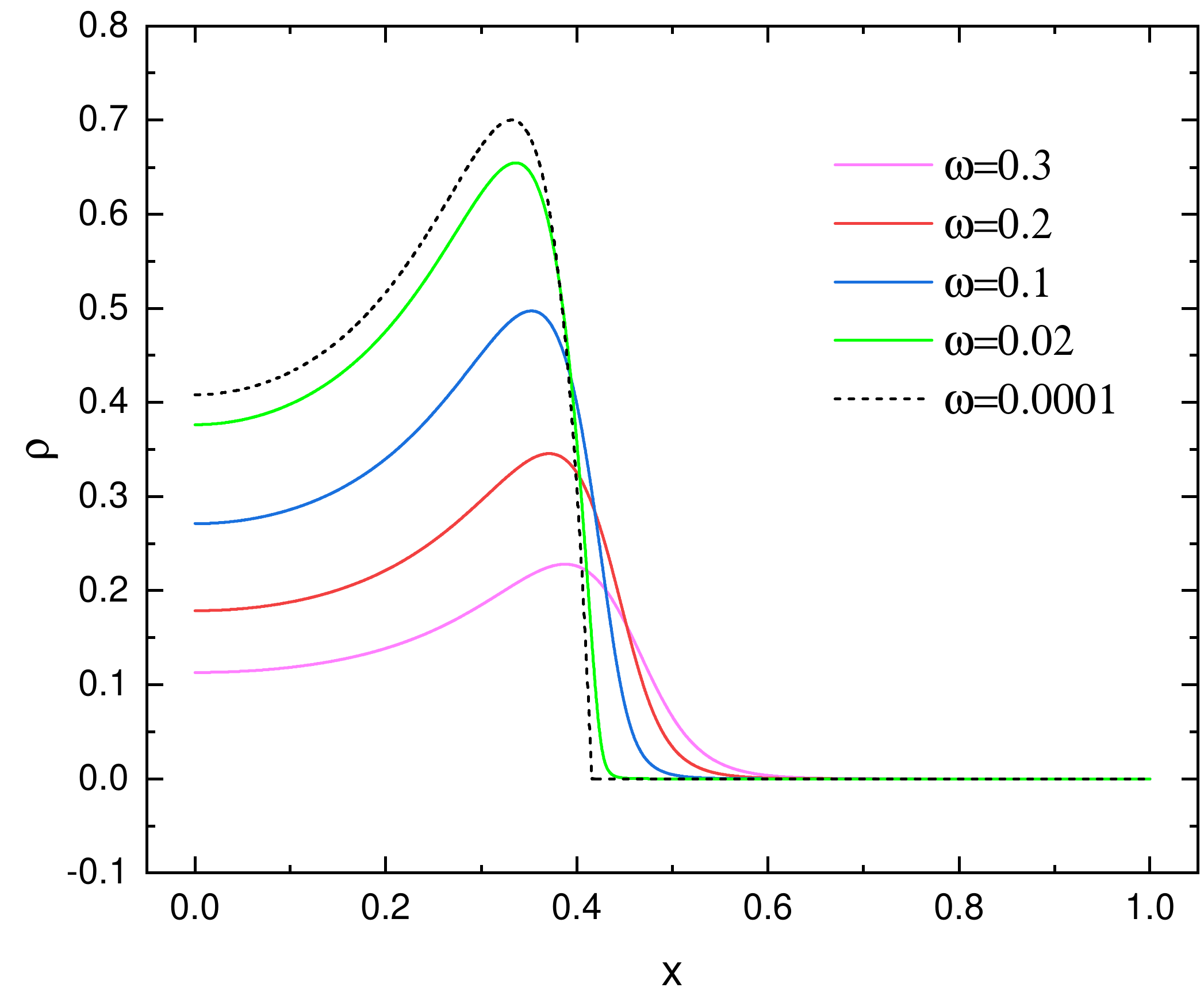}
  \end{center}
  \caption{The radial distribution of the  field functions with $q=0.65$.
  }\label{phase}
\end{figure}

\begin{figure}[]
  \begin{center}
  \includegraphics[width=9.8cm]{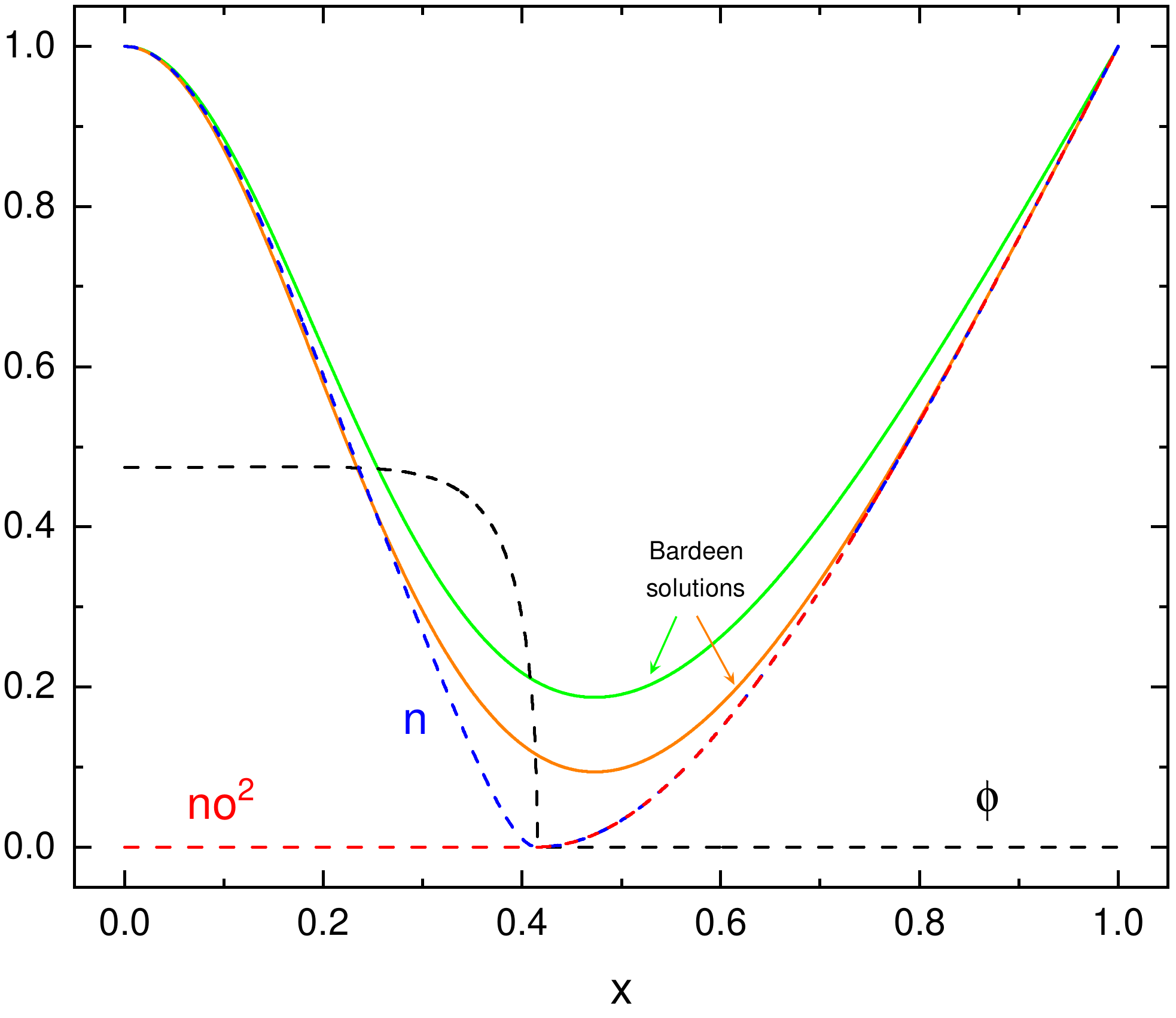}
  \end{center}
  \caption{The radial distribution   of the extreme solutions with $\omega \rightarrow  0 $. The green and orange curves represent the single Bardeen solutions with $s = 0.2$  and $s = 0.1794$, respectively. All curves have the same value of $q=0.65$.
  }\label{phase1}
\end{figure}
In this section, we present our numerical solutions.
For convenience, we set $s=0.2$ (except where specifically indicated).
First of all, in Fig. \ref{phase}, we give the configuration of the Bardeen-boson star solutions with
different values of the frequency $\omega$, in which the black dashed line  represents the case of approaching zero frequency.
The distribution of the function of $\phi$ is shown in the lower left panel,
we can see that the scalar field of these solutions converges at a certain radius $r_c$ in radial coordinate while rapidly decaying beyond that radius, forming a steep surface.
The component  $g_{tt}= n o^2$ of the metric in the upper right panel also approaches zero rapidly due to the influence of the matter field. As the frequency approaches zero, the $g_{tt}$ is nearly zero inside
 the region  $r<r_c$. In the lower right panel, it is shown that near  zero  frequency,  the energy density $\rho=T^{(2)0}_{~~\;0}$ of the scalar matter field,
has a very steep wall in the vicinity of the inner region near $r_c$.

To further understand the properties of these extreme solutions with $\omega \rightarrow  0 $, in Fig. \ref{phase1}
we introduce two different single Bardeen solutions with $s=0.2$ (green) and $s = 0.1794$ (orange), respectively. The dashed lines represent the  extreme solution with $\omega =0.0001 $ in Fig. \ref{phase}. All curves have the same value of $q=0.65$ and the orange and red dashed lines have the same mass. From the perspective of an observer at infinity, it is impossible to distinguish whether it is an equivalent Bardeen solution with orange dashed line or an extreme black hole with red  dashed line.
It is worth noting that, despite the fact that scalar fields are mostly distributed within the interior of the $r_c$ for the extreme solutions with $\omega \rightarrow  0 $, there is still a small amount distributed near $r>r_c$. In this case, the extreme Bardeen-boson solutions can be seen as a black hole with a few scalar hairs, which can be regarded as a form of ``soft hair" mentioned in \cite{Hawking:2016msc}.

\begin{figure}[]
  \begin{center}
  \includegraphics[width=8.1cm]{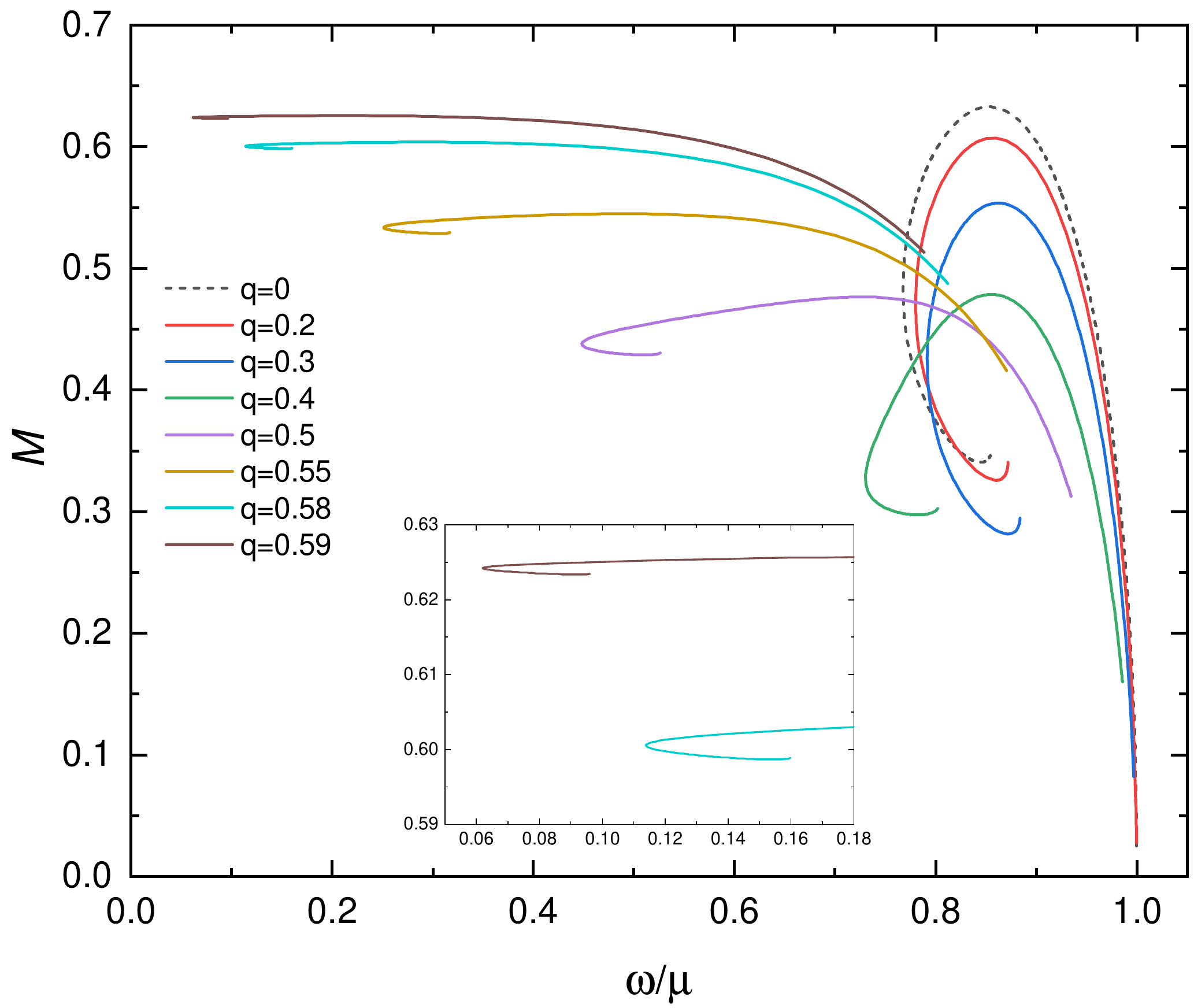}
   \includegraphics[width=8.2cm]{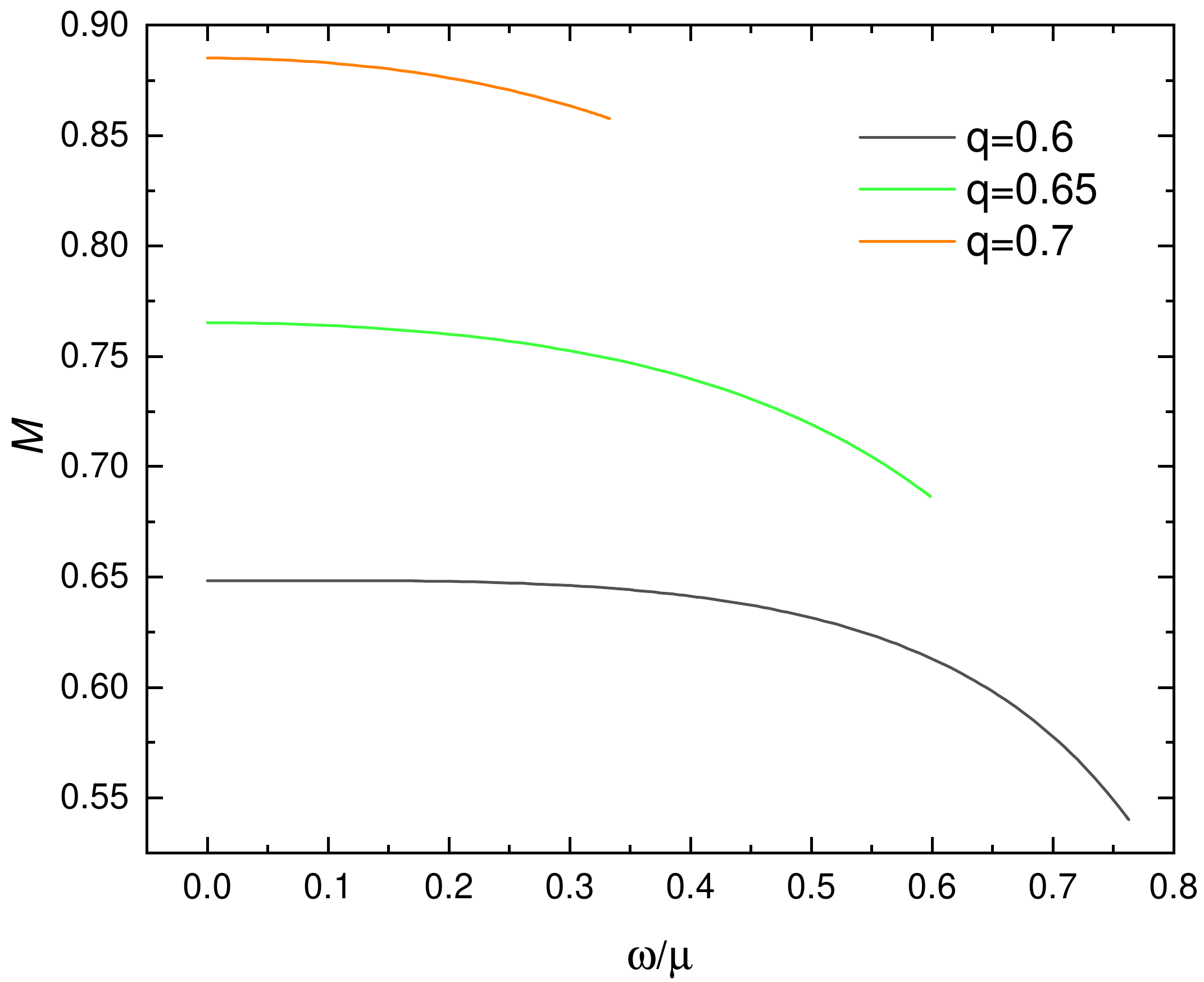}
     \includegraphics[width=8.2cm]{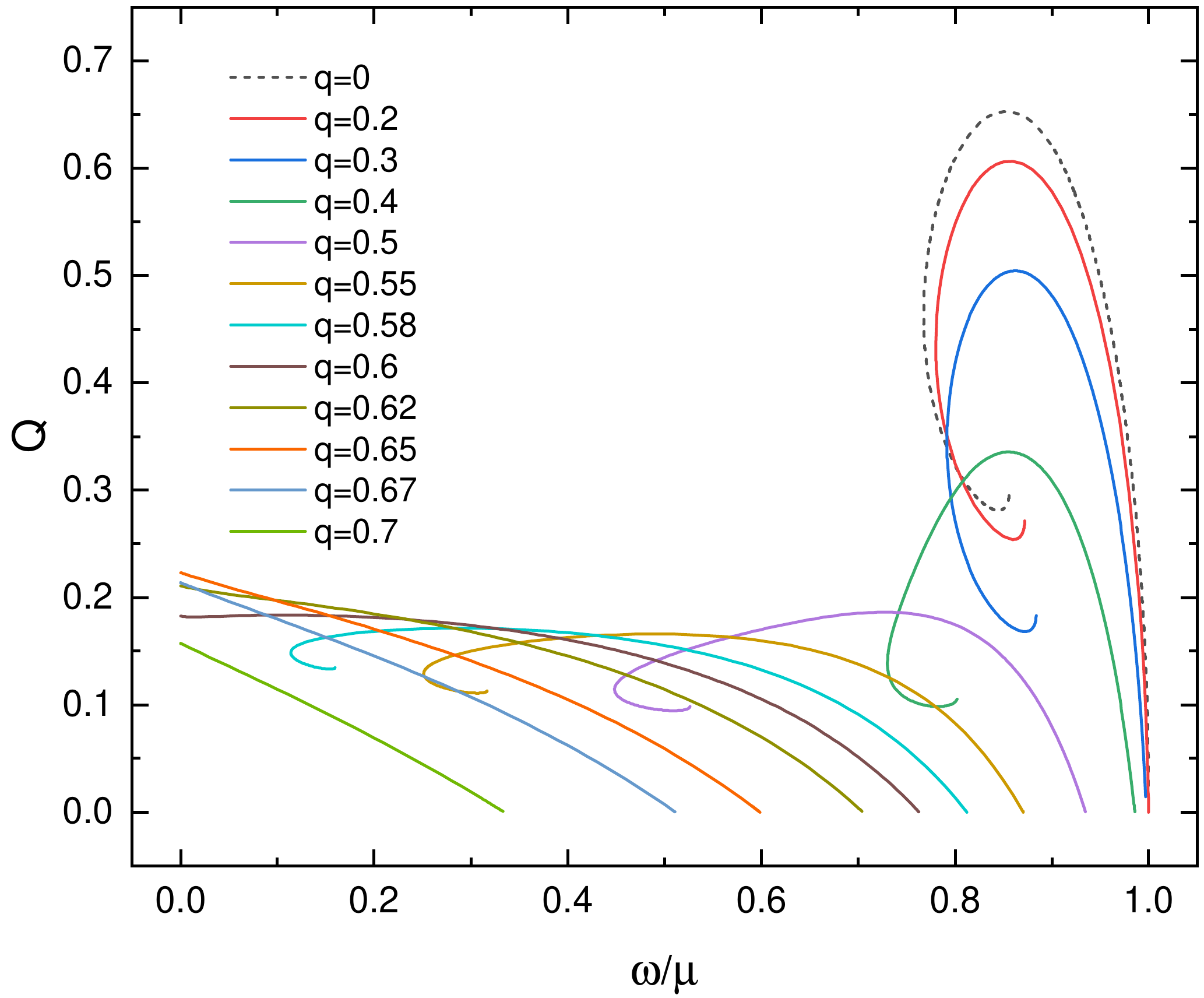}

  \end{center}
  \caption{The mass $M$ and Noether charges $Q$ of complex scalar field versus the frequency $\omega$ with the different values of magnetic charges $q$.
  }\label{phase4}
\end{figure}
In Fig. \ref{phase4}, we
exhibit the mass $M$ and Noether charges $Q$ of complex scalar field versus the frequency $\omega$ with the different values of magnetic charges $q$, the
black dashed curve corresponds to the single boson star solutions, in which the curve forms a spiral. Instead,
when $q$ starts to increase from $0$ value,
the curves start to deform and gradually unfold, and the range of the second branch becomes increasingly shorter. Then,  as $q$ continues to increase a critical value near $q=0.6$,
the second branch disappears, and the multi-valued curve will become a single-valued curve.

\section{Conclusions}
In this article, we investigated the model of Einstein-Klein-Gordon theory coupled to a nonlinear electrodynamics. When the scalar field vanishes, it corresponds to the well-known Bardeen black hole solution. However, we found that when the scalar field is present, we failed to find a black hole solution. Instead, we obtained a solution resembling a boson star, which lacks a event horizon. Under certain parameter values,
 the scalar field  converges at a certain radius $r_c$ in radial coordinate while rapidly decaying beyond that radius,
and the energy density of the scalar matter field has a very steep wall-liked distribution  in the vicinity of the inner region near $r_c$.
Outside this critical position,
 these boson star solutions tend to infinitely approach what is known as an extreme black hole. However, it is important to emphasize that these solutions did not possess an event horizon.

There are many interesting extensions of our work. First, in this work, we considered complex scalar fields as the matter content. However, it is also possible to replace the complex scalar field with other types of matter fields to verify if black holes without an event horizon exist as well.
Secondly, the study of Hawking radiation for such black holes is indeed worth investigating. Moreover, since there is no event horizon, the information loss paradox caused by Hawking radiation would not arise.
Furthermore, although our results are based on the Bardeen model, we believe similar results can be extended to other regular black hole models. Additionally, the stability of these Bardeen-boson stars is also an important aspect for further investigation.

\section{Acknowledgment}
I am immensely grateful for my  family. Your love, encouragement, and unwavering support have  made me strong and courageous, and has given me infinite motivation. Love you all!

\end{document}